\documentclass{article}
\usepackage{fortschritte}
                                            
\def\beq{\begin{equation}}  
\def\eeq{\end{equation}}   

\begin {document}                 
\def\titleline{Quantum Information and General Relativity}

\def\email_speaker{
{\tt peres@technion.ac.il}}
\def\authors{Asher Peres}
\def\addresses{Technion --- Israel Institute of Technology, Haifa}

\def\abstracttext{
The Einstein-Podolsky-Rosen paradox (1935) is reexamined in the light
of Shannon's information theory (1948). The EPR argument did not take
into account that the observers' information was localized, like any
other physical object. General relativity introduces new problems:
there are horizons which act as on-way membranes for the propagation of
quantum information, in particular black holes which act like sinks.}

\large
\makefront
\section{SPQR}
It is a pleasure to dedicate this article to Francesco DeMartini on the
occasion of his 70th birthday. Francesco is a magician of experimental
demonstrations for {\bf S}pin {\bf P}olarized {\bf Q}uantum {\bf
R}adiation. I am only a theorist, which is much easier. 

When I was a young man, my thesis adviser was Nathan Rosen, and
the subject was the existence of gravitational radiation in general
relativity. Only much later, I seriously learnt quantum mechanics,
and still much later information theory. I now want to return to my
roots and try to combine all these subjects together.

\section{The Einstein-Podolsky-Rosen dilemma}

Rosen had been a post-doc of Einstein at the Institute for Advanced
Studies in Princeton. One day, at the traditional 3~o'clock tea, as
Rosen mentioned to Einstein a fundamental issue of interpretation
related to entangled wave-functions, Einstein immediately saw the
implications for his long standing disagreement with Bohr. As they
discussed the problem, Boris Podolsky joined the conversation, and
later proposed to write an article. Einstein acquiesced. When he
later saw the text, he disliked the formal approach, but agreed to
its publication~\cite{epr}. Then, as soon as the EPR article appeared,
Podolsky relased its contents to the New York Times (4 May 1935,
page 11) in a way implying that the authors had found that quantum
mechanics was faulty. This infuriated Einstein, who after that no
longer spoke with Podolsky.

The EPR ``paradox'' drew immediate attention. Niels Bohr \cite{bohr}
found the reasoning faulty, because it contradicted his complementarity
principle. Bell, in his first article on hidden variables and
contextuality~\cite{bell66}, wrote ``the Einstein-Podolsky-Rosen
paradox is resolved in the way which Einstein would have liked least."
Actually, the example given by Bell in the proof of his celebrated
theorem~\cite{bell64} is based on a much simpler entangled system:
two spin-$1\over2$ particles in a singlet state~\cite{bohm}.

The EPR article was not wrong, but it had been written too early.
Only some years later, in 1948, Claude Shannon published his theory
of information~\cite{shannon} (and it took many more years before
the latter was included in the physicist's toolbox). Shannon was
employed by the Bell Telephone Company and his problem was to make
communication more efficient. Shannon showed that information could be
given a quantitative measure, that he called {\it entropy\/}. It was
later proved that Shannon's entropy is fully equivalent to ordinary
thermodynamical entropy~\cite{bennett}. Information can be converted
to heat and can perform work. Information is not just an abstract
notion~\cite{landauer}. It requires a physical carrier, and the latter
is (approximately) {\it localized\/}. After all, it was the business of
the Bell Telephone Company to transport information from one telephone
to another telephone, in a different location.

In the EPR article, the authors complain that ``it is possible to assign
two different wave functions to \ldots\ the second system,'' and then,
in the penultimate paragraph, they use the word {\it simultaneous\/} no
less than four times, a surprising expression for people who knew very
well that this term was undefined in the theory of relativity. Let
us examine this issue with Bohm's singlet model. One observer,
conventionally called Alice, measures the $z$-component of the spin of
her particle and finds $+\hbar/2$. Then she {\it immediately\/} knows
that if another distant observer, Bob, measures (or has measured,
or will measure) the $z$-component of the spin of his particle, the
result is certainly $-\hbar/2$. One can then ask: when does Bob's
particle acquire the state with $s_z=-\hbar/2$?

This question has two answers. The first answer is that the question is
meaningless --- this is undoubtedly true. The second answer is that,
although the question is meaningless, it has a definite answer: Bob's
particle acquires this state {\it instantaneously\/}. This then raises a
new question: in which Lorentz frame is it instantaneous? Here, there is
also a definite answer: it is instantaneous in the Lorentz frame that we
arbitrarily choose to perform our calculations~\cite{interv2}. Lorentz
frames are not material objects: they exist only in our imagination.

When Alice measures her spin, the information she gets is localized at
her position, and will remain so until she decides to broadcast it.
Absolutely {\it nothing\/} happens at Bob's location. From Bob's point
of view, all spin directions are equally probable, as can be verified
experimentally by repeating the experiment many times with a large
number of singlets without taking in consideration Alice's results.
Thus, after each one of her measurements, Alice assigns a definite
pure state to Bob's particle, while from Bob's point of view the state
is completely random ($\rho$ is proportional to the unit matrix). It
is only if and when Alice informs Bob of the result she got (by mail,
telephone, radio, or by means of any other material carrier, which is
naturally restricted to the speed of light) that Bob realizes that his
particle has a definite pure state. Until then, the two observers can
legitimately ascribe different quantum states to the same system. For
Bob, the state of his particle suddenly changes, not because anything
happens to that particle, but because Bob receives information about
a distant event. Quantum states are not physical objects: they exist
only in our imagination.

\section{Curved Spacetime}

It took Einstein more than ten years of intensive work to progress
from special relativity to general relativity. Despite its name, the
latter is not a generalization of the special theory, but a radically
different construct: spacetime is not only a passive arena where
dynamical processes take place, but has itself a dynamical nature. At
this time, there is no satisfactory quantum theory of gravitation (in
spite of seventy years of efforts by leading theoretical physicists).

Concepts of quantum information were recently invoked in several
problems of quantum gravity and quantum cosmology. Even if we still
consider spacetime as a passive arena endowed with a Riemannian metric,
instead of the Minkowski metric of special relativity, the difference
between them is essential: it is necessary to introduce notions of
topology, because it may be impossible to find a single coordinate
system that covers all of spacetime. To achieve that result, it may be
necessary to use several coordinate patches, sewed to each other at
their boundaries. Then in each patch, the metric is not geodesically
complete: a geodesic line stops after a finite length, although there
is no singularity there. The presence of singularities (points of
infinite curvature) is another consequence of Einstein's gravitational
equations. It is likely that these equations, which were derived
and tested for the case of moderate curvature, are no longer valid
under such extreme conditions. I shall not speculate on this issue,
and shall restrict my attention to the behavior of quantum systems
in the presence of {\it horizons\/}, in particular of black holes
that result from concentrations of matter so large that their gravity
pull prevents the escape of light. In other words, a future horizon
is formed.  While Unruh's horizons~\cite{unruh} appear for uniformy
accelerated observers whose asymptotic speed approaches $c$, a black
hole horizon affects every observer.

Even in classical general relativity, there is a serious
difficulty with the second law of thermodynamics when a black hole
is present: if we drop ordinary matter into a black hole, it will
disappear into a spacetime singularity, together with its entropy
$S$. No compensating gain of entropy occurs, so that the total
entropy in the universe decreases.  One could attempt to salvage
the second law by invoking the bookkeeping rule that one must
continue to count the entropy of matter dropped into a black hole
as still contributing to the total entropy of the universe.
However, the second law would then be observationally
unverifiable.

Suppose now that the matter that has fallen inside the horizon had
quantum correlations with matter that remained outside. How is
such a state described by quantum theory? Are these correlations
observable? This problem is not yet fully understood, although
such correlations play an essential role in giving to Hawking
radiation a nearly exact thermal character~\cite{wald}. It is
hard to imagine a mechanism for restoring the correlations during
the process of black hole evaporation. On the other hand, if the
correlations between the inside and the outside of a black hole
are not restored during the evaporation process, then by the time
that the black hole has evaporated completely, an initial pure
state will have evolved to a mixed state, and some ``information''
will have been lost.

It has often been asserted that the evolution of an initial pure
state into a final mixed state conflicts with quantum mechanics, and
this issue is usually referred to as the ``black hole information
loss paradox.''  These pessimistic views are groundless. When
black hole thermodynamics appeared in the 70's, notions such as
POVMs and completely positive maps were unknown to the relativistic
community. Today, we know that the evolution of pure states into
mixtures is the general rule when a classical intervention is imposed
on a quantum system. However, this issue may be conclusively resolved
only after there is a consistent theory of quantum gravity, allowing
meanwhile for a number of tantalizing speculations. Here we present some
of the most popular alternatives of what happens with the ``information"
when a black hole evaporates.

\begin{itemize}

\item Information is lost: this is a fundamental feature of quantum
theory in the presence of black holes and not just an effective
description.

\item There is no information loss: if the spectrum of Hawking
radiation is analyzed carefully, there may be enough non-thermal
features to encode all the information. Hod~\cite{hod} estimated
that, under suitable assumptions about black hole quantization, the
maximal information emission rate may be sufficient to recover all
the information from the resulting discrete spectrum of the radiation.

\item Information comes out at the end, at the Plank scale physics.
There is a stable black hole remnant with about the Planck mass
(0.02 $\mu$g) and information is somehow encoded in it~\cite{acn}.
\end{itemize}

Still a different scenario is implied by the works of
Gerlach~\cite{gerlach} and Boulware~\cite{boulware}: a particle that
falls into an eternal black hole crosses the horizon after an infinite
amount of the coordinate time $t$, but only a finite amount of its own
proper time. On the other hand, the evaporation of a black hole takes
a finite amount of the coordinate time, which is the physical time of a
distant observer. From the point of view of the infalling observer, the
horizon always appears to recede before her, until it finally disappears
(or shrinks to the Planck scale) and the region ``beyond the horizon''
is unattainable. The distant observer sees the infalling one quickly
arrive arbitrarily close to the effective horizon, then she is nearly
``frozen" there for an exceedingly long time, and finally either the
black hole evaporates or the universe collapses.\medskip

{\bf Acknowledgement.} This work was supported by the Gerard Swope Fund.

\end{document}